\documentclass[aps,pra,reprint,onecolumn,groupedaddress]{revtex4-2}
\usepackage[hidelinks]{hyperref}

\usepackage{graphicx}
\usepackage{color}
\usepackage{amsmath}
\usepackage{amsfonts}
\usepackage{amssymb}
\usepackage{float}

\begin{document}
\title{Fast classical simulation of evidence for the utility of quantum computing before fault tolerance}
\author{Tomislav Begu\v{s}i\'{c}}
\author{Garnet Kin-Lic Chan}
\email{gkc1000@gmail.com}
\affiliation{Division of Chemistry and Chemical Engineering, California Institute of Technology, Pasadena, California 91125, USA}
\date{\today}

\begin{abstract}
We show that a classical algorithm based on sparse Pauli dynamics
can efficiently simulate quantum circuits studied in a recent experiment on 127 qubits of IBM's Eagle processor [\textit{Nature} \textbf{618}, 500 (2023)]. Our classical simulations on a single core of a laptop are orders of magnitude faster than the reported walltime of the quantum simulations, as well as faster than the estimated quantum hardware runtime without classical processing, and are in good agreement with the zero-noise extrapolated experimental results.
\end{abstract}

\maketitle

Clifford-based methods \cite{Bravyi_Gosset:2016,Bravyi_Howard:2019,Bennink_Pooser:2017,Qassim_Emerson:2019,begusic2023simulating} are a promising approach for the
classical simulation of quantum circuits. 
The complexity of these approaches is primarily governed by
the number of non-Clifford gates because they use the property that Clifford gates, which include all Pauli rotation gates with angles $\theta = k\pi/2$ (for $k$ integer), can be simulated efficiently in polynomial time. In contrast to other common techniques, such as tensor network methods~\cite{gray2021hyper,huang2021efficient,gray2022hyper}, Clifford-based approaches are not limited by the amount of entanglement generated by the quantum circuit. 

A recent experiment presented evidence for the utility of quantum computing before fault tolerance by comparing results from the zero-noise extrapolated \cite{temme2017error} quantum simulation of the kicked Ising model on up to 127 qubits to those from classical matrix product state (MPS) and isometric tensor network state (isoTNS) simulations~\cite{kim2023evidence}. 
Ref.~\cite{kim2023evidence} showed that these tensor network simulations
failed to produce sufficiently accurate results for the set of considered circuits, even in cases when all the rotation gates were Clifford or near-Clifford. At the same time, the number of non-Clifford gates in the experiment prohibit the use of exact exponentially-scaling Clifford approaches~\cite{Bravyi_Gosset:2016,Bravyi_Howard:2019,Bennink_Pooser:2017,Qassim_Emerson:2019}. Here we apply a variant of a recently introduced Clifford perturbation theory \cite{begusic2023simulating}, which we term sparse Pauli dynamics (SPD), to the simulation of circuits of Ref.~\cite{kim2023evidence}. The technique allows us to truncate the exponential growing set of Pauli operators generated by the non-Clifford gates, yielding an approximation algorithm that takes advantage of near-Clifford structure.

\begin{figure}
    \centering
    \includegraphics[width = 0.32\textwidth]{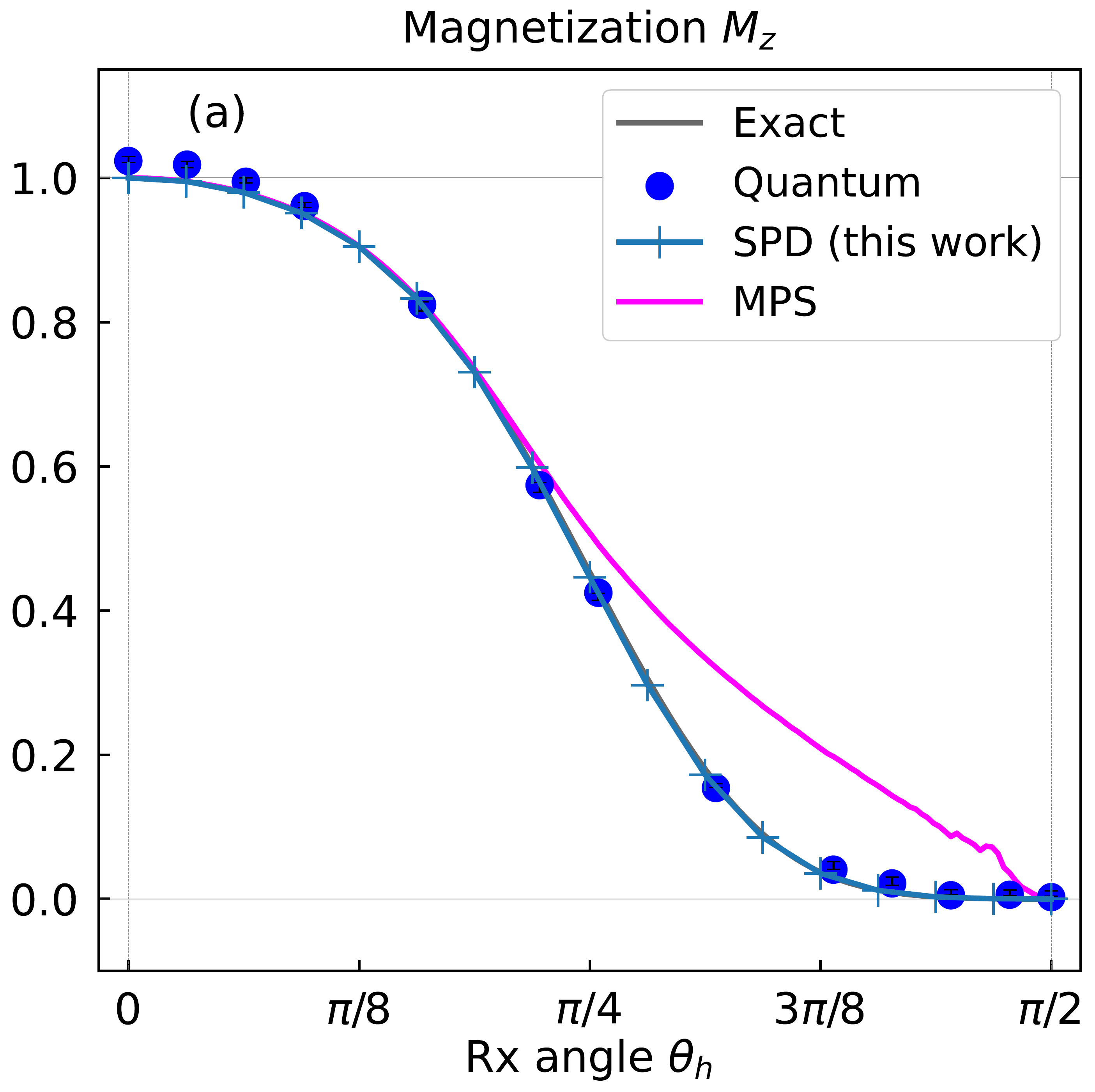}
    \includegraphics[width = 0.32\textwidth]{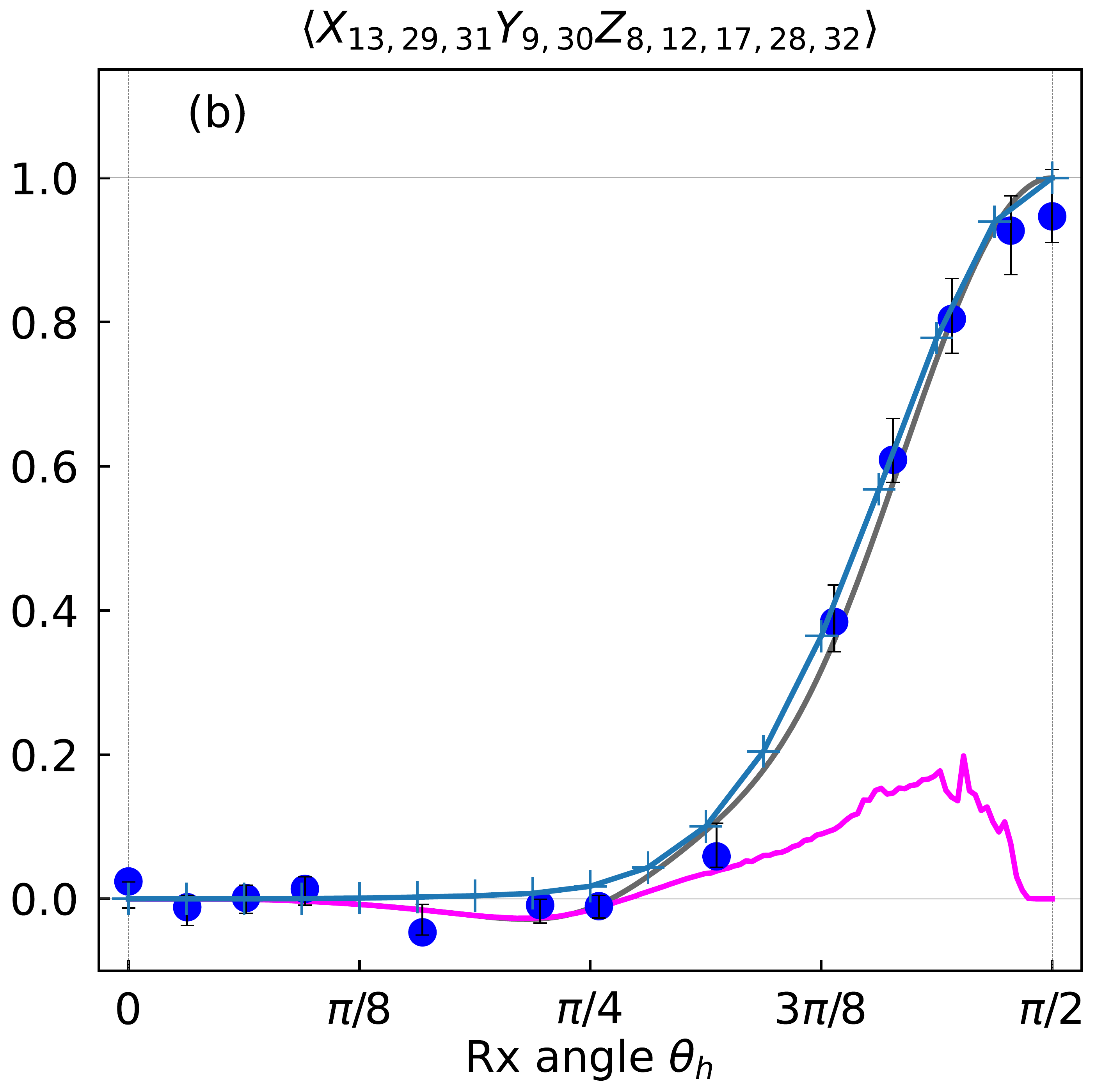}
    \includegraphics[width = 0.32\textwidth]{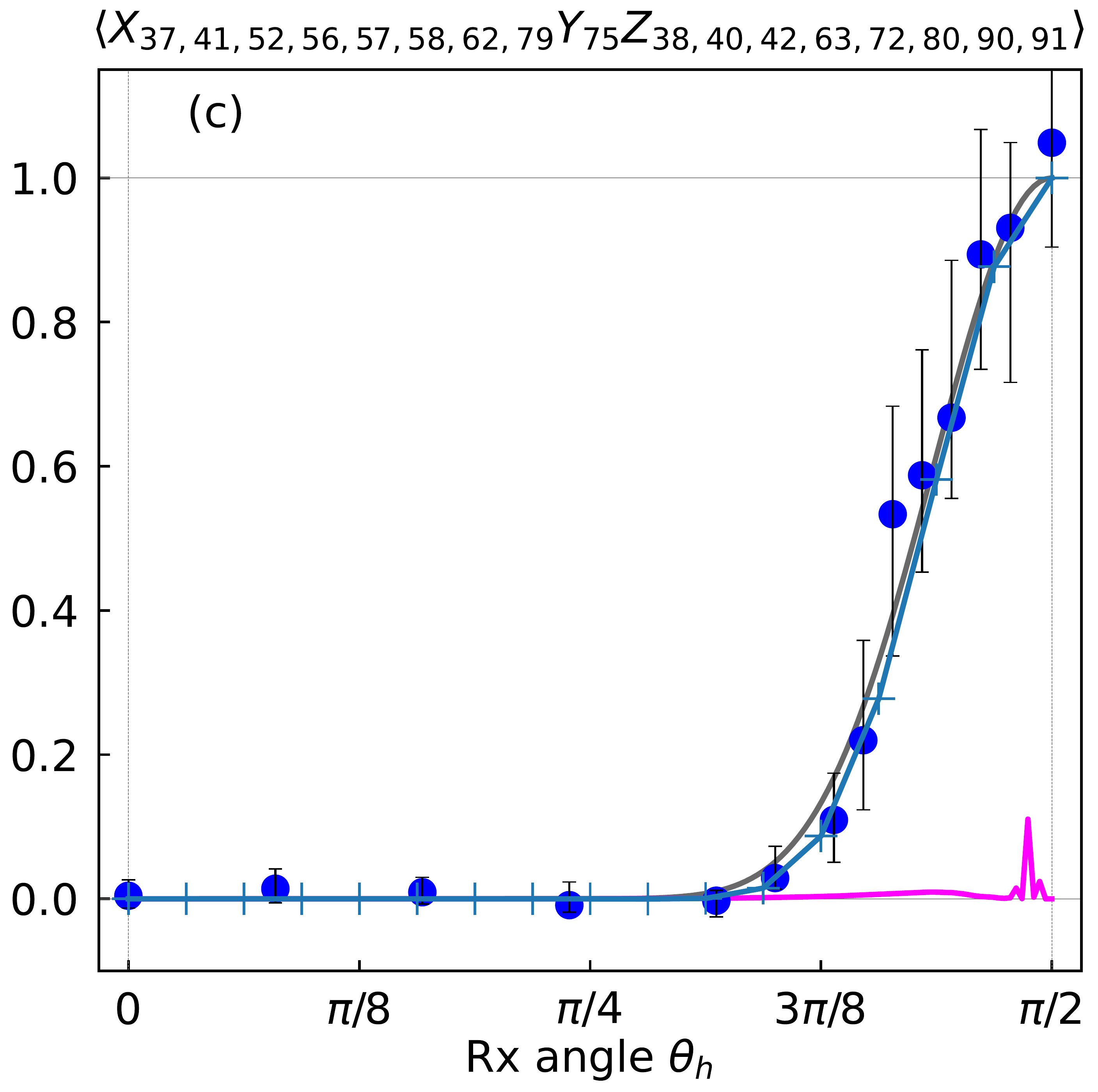}
    \includegraphics[width = 0.32\textwidth]{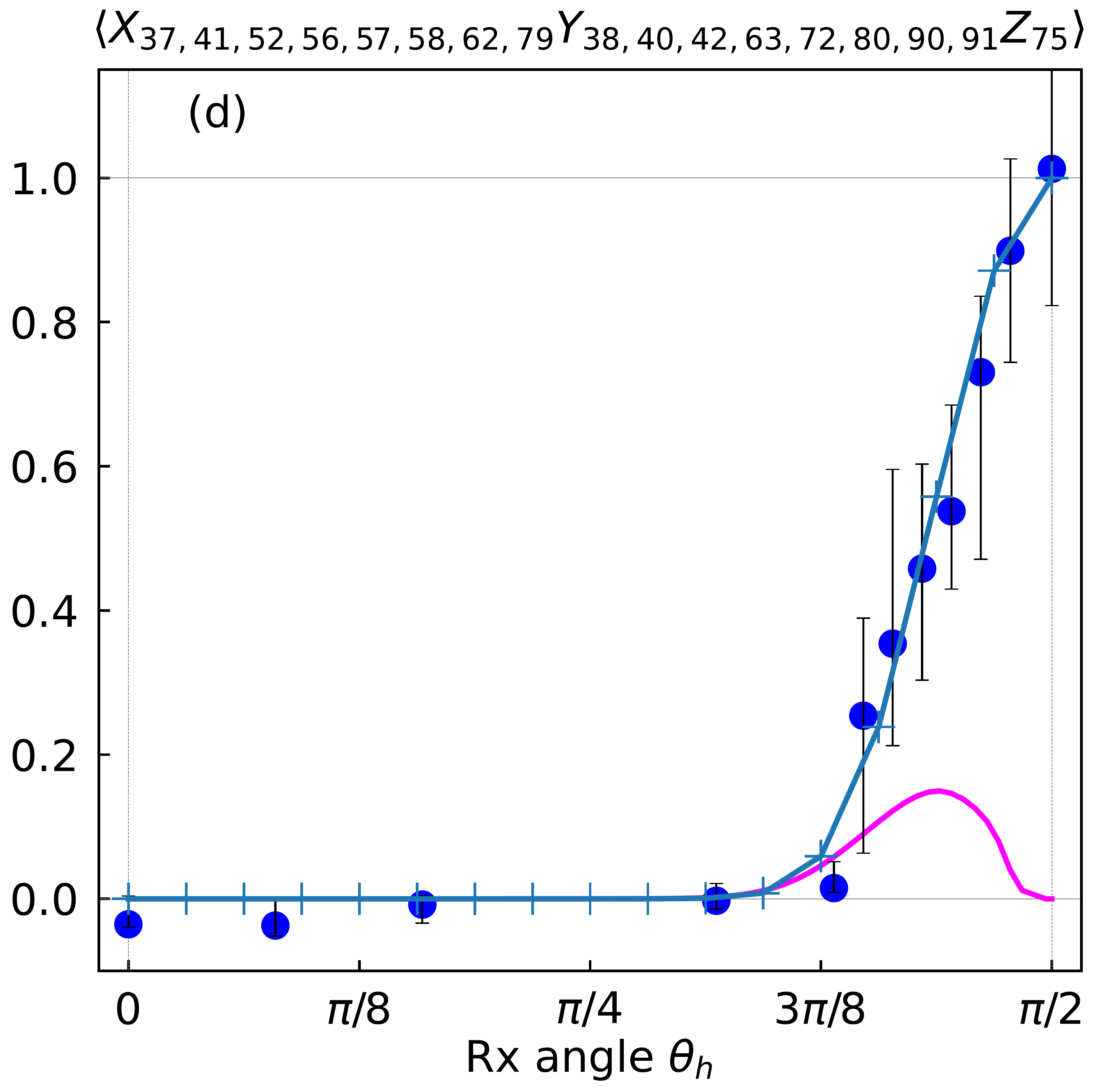}
    \includegraphics[width = 0.32\textwidth]{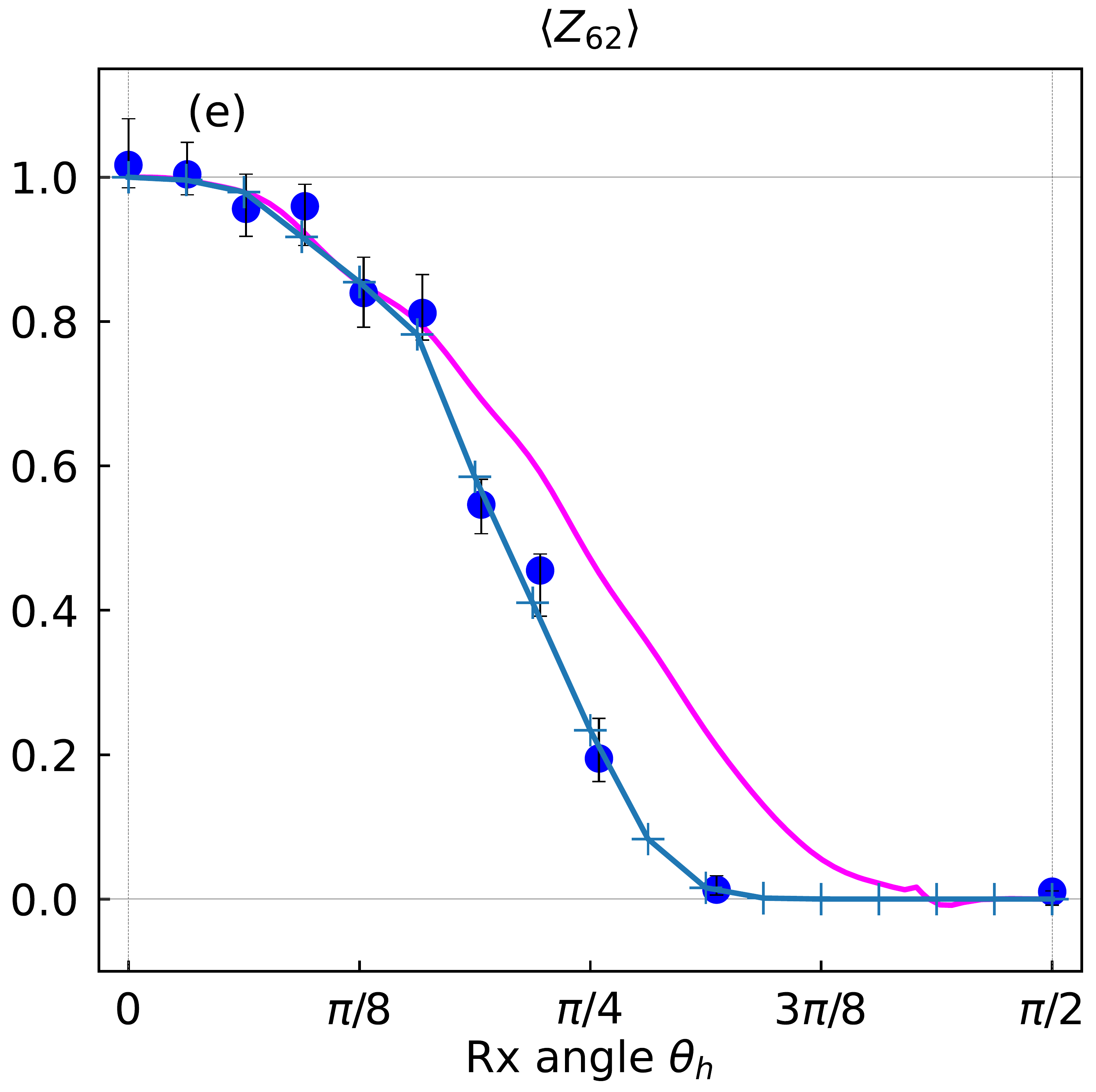}
    \caption{Expectation values of different observables considered in Ref.~\cite{kim2023evidence}. Exact results [for panels (a)--(c)], quantum simulation experiment (``Quantum''), and MPS results are reproduced from Figs.~(3a)--(4b) of Ref.~\cite{kim2023evidence}. Following that reference, the magnetization $M_z = \sum_q Z_q / n$ (a), weight-10 (b), and weight-17 (c) observables are simulated for a quantum circuit consisting of five Trotter steps (amounting to five layers of non-Clifford gates), another weight-17 observable (d) is computed for a quantum circuit consisting of five Trotter steps and an additional $R_X$ layer (six non-Clifford layers in total), and $\langle Z_{62} \rangle$ in (e) is simulated for a circuit composed of twenty Trotter steps (twenty non-Clifford layers). For visual simplicity, values in (c) and (d) were negated. The SPD results were computed at $\theta_h = k \pi / 32$ for integer $k$ between 0 and 16 (``+'' markers).}
    \label{fig:main}
\end{figure}

\

\noindent\textbf{Results.} Following Ref.~\cite{kim2023evidence}, we consider a quantum circuit composed of alternating layers $R_{ZZ} = \prod_{\langle i,j \rangle}\exp(i \frac{\pi}{2} Z_i Z_j)$ and $R_X(\theta_h) = \prod_i \exp(-i \frac{\theta_h}{2} X_i)$ of Pauli rotation gates, i.e.,
\begin{equation}
    U = [R_{ZZ} R_X(\theta_h)]^t,
\end{equation}
where $\langle i, j\rangle$ denotes that $i$ and $j$ qubits are neighbors in the heavy-hexagon lattice of IBM's Eagle processor \cite{kim2023evidence} and $t$ is the number of Trotter steps. Since the entangling gates in $R_{ZZ}$ are Clifford, we first transform the circuit to consist only of non-Clifford Pauli rotation gates $R_{\tilde{X}^{(m)}}$, where $\tilde{X}_i^{(m)} = (R_{ZZ}^{\dagger})^m X_i R_{ZZ}^m$. We can further perform an angle transformation of non-Clifford rotation gates $R_X(\theta_h) = R_X(\tilde{\theta}_h) R_X(\pi/2)$. We apply this transformation in the simulation of weight-10 and weight-17 observables (see below), which under this transformation convert into single-qubit Pauli operators. To handle the remaining non-Clifford gates, we then perform the sparse Pauli dynamics, which works in the Heisenberg picture of the observable in the Pauli representation (see Method). Its accuracy is controlled by a truncation order $K$, which limits the size of the Pauli basis.

Figure~\ref{fig:main} compares our classical simulation results with those obtained in Ref.~\cite{kim2023evidence} by a numerically exact method [for panels (a)--(c)], the quantum simulation experiments, and the MPS approach. The maximum truncation orders were $K=10$ (a, b, e) and $K=6$ (c, d). With these parameters, each point in Fig.~\ref{fig:main} takes between one and two minutes on a single processor of a laptop computer, which is two orders of magnitude faster than the reported quantum wall-clock run time for Figs.~\ref{fig:main}d,~e (4 h and 9.5 h, respectively) and faster than the hypothesized run time of  IBM's Eagle processor {without} any classical processing steps (estimated at about 5~min for Fig.~\ref{fig:main}e). Although 
the chosen truncation orders do not yield 
numerically exact results (which can be easily verified by the small deviations in panels (a)--(c)) and more detailed studies of convergence should be carried out, the results are comparable to (and generally {well within}) the estimated errors of the quantum simulation experiment and much more accurate than the MPS results of Ref.~\cite{kim2023evidence}.

Overall, our results demonstrate that our simple sparse Pauli dynamics approach can faithfully simulate the quantum experiments recently performed on the new Eagle processor of IBM. Two days before this preprint was posted, a preprint was posted by Tindall et al. \cite{tindall2023efficient} that showed that an alternative tensor network method that takes advantage of the qubit connectivity of the device also outperforms the MPS or isoTNS~\cite{zaletel2020isometric} methods employed in Ref.~\cite{kim2023evidence} and can also simulate the quantum experiments. In contrast to that work, our method is not specially constructed for a given qubit connectivity map, nor does it rely on low entanglement. 
That both classical approaches succeed illustrates the rich landscape of approximate classical algorithms that remain to be explored. 
Finally, we believe that the method we describe here holds promise not only for quantum circuit simulations, but also for more general simulation problems in quantum dynamics.

\

\noindent\textbf{Method.} Consider the unitary composed of Pauli rotation gates after the Clifford transformation described above: 
$ U = \prod_i U_j(\theta_j) = \prod_j \exp(-i \theta_j P_j / 2)$, defined by the rotation angles $\theta_j$ and Hermitian Pauli operators $P_j$. (Note that because the Clifford gates have been applied, the Paulis $P_j$ are not those of the original circuit, and all the rotations are non-Clifford). 

We compute the expectation value of a Pauli observable
\begin{equation}
    \langle O \rangle = \langle 0^{\otimes n} | U^{\dagger} O U |0^{\otimes n} \rangle \label{eq:exp_val}
\end{equation}
by performing Heisenberg evolution of the observable under Pauli rotation gates, using
\begin{equation}
    U_j(\theta_j)^{\dagger} \sigma U_j(\theta_j) = \cos(\theta_j) \sigma + i \sin(\theta_j) P_j \sigma \label{eq:obs_rot}
\end{equation}
whenever $\{\sigma, P_j\}=0$ and $U_j(\theta_j)^{\dagger} \sigma U_j(\theta_j) = \sigma$ otherwise.
For Clifford gates, Eq.~(\ref{eq:obs_rot}) simply transforms a Pauli operator $\sigma$ into another Pauli operator, whereas for non-Clifford gates it results in a sum of two Pauli operators, leading to an exponential computational cost. This sum can be written as
\begin{align}
\langle O \rangle &= \sum_{k=0}^{N} E^{(k)}, \label{eq:O_sum}\\
E^{(k)} &= i^k \sum_{1 \leq j_1 < j_2 < \cdots < j_k \leq N} c_{j} \sin(\theta_{j_1}) \cdots \sin(\theta_{j_k}) \langle 0^{\otimes n}| P_{j_k} \cdots P_{j_1}O |0^{\otimes n} \rangle,\label{eq:pert_k}
\end{align}
where $N$ is the number of gates, $j_1, \dots, j_k$ are selected so that $\{ P_{j_m}, P_{j_{m-1}} \cdots P_{j_{1}} O \} = 0$, and $c_j$ is a product of cosines of $\theta_{l \notin \{j_1, \dots, j_k\}}$ \cite{begusic2023simulating}. To truncate the exponentially growing sum of Pauli operators, we recently introduced a perturbation-inspired expansion \cite{begusic2023simulating}, which assumes sufficiently small angles $\theta_j$ [i.e., small $\sin(\theta_{j_m})$] that suppress high-order terms $E^{(k)}$. This assumption is supported by the fact that the rotation gates can be rewritten as $U_j(\theta_j) = U_j(\tilde{\theta}_j)U_j(k\pi/2)$ to ensure that $\tilde{\theta}_j \in [-\pi/4, \pi/4]$, while $U_j(k\pi/2)$ are Clifford and do not directly contribute to the computational cost.

In contrast to our previous work, here we do not directly truncate the series (\ref{eq:O_sum}) up to a specified order $K$ but use the same order concept to adaptively increase the Hilbert (operator) subspace on which the approximate Heisenberg-evolved operator is supported. Because the Heisenberg observable is supported in a sparse subspace of the Pauli basis, we refer to this scheme as sparse Pauli dynamics. For related concepts, see also~\cite{kuprov2007polynomially,rakovszky2022dissipation}.
Specifically, let
\begin{equation}
    O_j = \sum_{\sigma \in \mathcal{P}_j} a_{\sigma, j} \sigma
\end{equation}
be the observable evolved with $j$ gates, $\mathcal{P}_j$ a set of Pauli operators at step $j$, and $a_{\sigma, j}$ complex coefficients. In addition, we assign an order $k_{\sigma}$ to each Pauli operator $\sigma$, defined as the smallest number $k$ for which $\sigma = P_{j_k} \cdots P_{j_1} O$ (note that $P_{j_i}$ is not in general a single qubit Pauli, so the weight of the operator string is in general $> k+1$). The rotation gate $U_{j+1}(\theta_{j+1})$ will commute with a subset of Pauli operators in $\mathcal{P}_j$ and will transform the rest according to Eq.~(\ref{eq:obs_rot}). Then, the coefficients are updated as
\begin{equation}
    a_{\sigma, j+1} = \cos^{\xi_{\sigma, P_{j+1}}}(\theta_{j+1}) a_{\sigma, j} + i \xi_{\sigma, P_{j+1}} \sin(\theta_{j+1}) a_{P_{j+1}\sigma, j}, \label{eq:coeff_update}
\end{equation}
where $\xi_{\sigma, \sigma^{\prime}}$ is 1 if the operators anticommute and 0 otherwise. For compactness, in Eq.~(\ref{eq:coeff_update}) it is understood that $a_{\sigma, j} = 0$ if $\sigma \notin \mathcal{P}_j$, even though in practice this zero coefficient is not stored. The number of Pauli operators increases by one whenever $\xi_{\sigma, P_{j+1}} = 1$, $\sigma \notin \mathcal{P}_j$, and $\sigma^{\prime} = P_{j+1} \sigma \in \mathcal{P}_j$, because then we must add $\sigma$ in the set $\mathcal{P}_{j+1}$. To control the number of Pauli operators from growing exponentially, we discard $\sigma$ with $k_{\sigma} > K$, where $K$ is the maximum truncation order and a parameter controlling the accuracy of the simulation. Finally, the expectation value is obtained as
\begin{equation}
    \langle 0^{\otimes n} | O_N | 0^{\otimes n} \rangle = \sum_{\sigma \in \mathcal{P}_N} a_{\sigma, N} \langle 0^{\otimes n} | \sigma | 0^{\otimes n} \rangle.
\end{equation}
A Python implementation using IBM's Qiskit \cite{Qiskit} was employed in producing the results for Fig.~\ref{fig:main}.

\

\noindent\textbf{Acknowledgments.} TB and GKC were supported by the US Department of Energy, Office of Science, Office of Advanced Scientific Computing Research and Office of Basic Energy Sciences, Scientific Discovery through Advanced Computing (SciDAC) program under Award Number DE-SC0022088. TB acknowledges financial support from the Swiss National Science Foundation through the Postdoc Mobility Fellowship (grant number P500PN-214214). GKC is a Simons Investigator in Physics.

\bibliographystyle{aipnum4-2}
\bibliography{bibliography}

\end{document}